\documentstyle[art12]{article}
\title{The Problem of the Double Time-Ordered Operator Product
Calculation in the Nonequilibrium Quantum Field Theory}
\author{Tengiz M.~Bibilashvili\\ \it Institute of Physics, \\
\it Academy of Sciences of the Georgian Republic,\\ \it Tamarashvili 6,
Tbilisi, Georgian Republic, GE-380077\\ \it E-mail: tbib@physics.iberiapac.ge}
\date{}
\begin{document}
\maketitle
\begin{abstract}
It is shown that calculation of the operator's groups product with the
independent time ordering procedure in each one is important for some problems
in the nonequilibrium quantum field theory at finite temperature. This problem
is solved in terms of path integrals by means of time-integration contour
transformation. A role of thermal ghosts for this problem is also discussed.
\end{abstract}
\vspace{8cm}
{\bf IP HE-94/17}\\
{\it hep-ph 9412288}
\newpage
\section{Introduction}
\label{In}

Recently there has been much interest in the collective properties of particle
equilibrium \cite{NS,UMT} and nonequilibrium \cite{CH,RJ} systems in study of
the early Universe \cite{KSW,TK,HB} and of the heavy ion collisions \cite{JK}.
Both kind of systems are intrinsically nonequilibrium in nature, and only some
details of their behaviour may be considered in the frame of equilibrium
description.

The most useful in the quantum field theory (QFT) is the Green's function
method.  It was used by Schwinger, Keldysh \cite{SK} and others (see
\cite{JK,LvW,RR,HK} for a review) for the nonequilibrium case when the system
is
subjected to the ``mechanical'' perturbations (perturbations that may be
considered as a part of the Hamiltonian \cite{K}).  Of course, mechanical
perturbations produce ``thermal'' ones (variations of the temperature and other
thermodynamical parameters \cite{K}).  This property was used by many authors
\cite{OA} to estimate kinetic coefficients connected with thermal perturbations
(like thermal conductivity, shear and bulk viscosity etc.  \cite{GLW}).
Thermal
perturbations were taken into account explicitly by McLennan \cite{Mp,Mb} and
Zubarev \cite{Zub1,Zub}.  Temperature (and/or chemical potential etc.)  is
considered in mentioned approaches as some background field and is not
subjected
to any influence.  This theories are formulated in terms of density matrix and
they generalise the Gibbs method to the nonequilibrium case.  Similar but more
complicate from the technical point of view approaches were developed by some
other authors \cite{MG}.  Another formalism of the QFT systems description is
thermo-field dynamics constructed by Umezawa and co-authors \cite{UMT}.  It
coincides with other approaches in the stationary case but differs in the
nonequilibrium situation \cite{U}.

This paper is based on the Zubarev's density matrix, that is called by author
Nonequilibrium Statistical Operator (NSO) \cite{Zub}.  It was used previously
in
QFT (see \cite{HS,SS}) in the imaginary-time Matsubara formalism \cite{Ma}.
Our
consideration\cite{BP,B} is done in the real-time and in terms of path
integrals
\cite{NS}.  Consideration of the nonequilibrium is more natural in the
real-time
approach without procedure of analytic continuation.  It is shown in \cite{B}
that an account of the thermal perturbations leads to some theoretical
problems.
A calculation of the following unusual operator product is needed
\begin{equation} {\rm Tr}\left\{ {\rm
T}\left[\hat{\phi}(1)\cdots\hat{\phi}(n)\right] {\rm
T}\left[\hat{\phi}(n+1)\cdots \hat{\phi}(m)\right]\cdots\right\},
\label{1}
\end{equation}
where $\phi(i)=\phi(x_i)$ and ${\rm T}$ means time ordering of the operators.
The rarity of this expression is due to the presence of two ${\rm T}$-ordering
procedures.  It may be called ``double (multi) time-ordered product''.  I shall
solve this problem in terms of path integrals for quantum field theory in
vacuum
(Sec.  3) and in the heat bath (Sec.  4).

\section{How double perturbative product appear in the theory} \label{Ho}

In this section it is briefly reviewed how double-time operator product
(\ref{1}) occur in the nonequilibrium QFT \cite{B}.

An average value of some operator $\hat{A}$ may  be obtained  by the
following manner
\begin{equation}
<\hat{A}>={\rm Tr}\left\{\hat{\rho}\hat{A}\right\},
\label{2}
\end{equation}
where $\hat{\rho}$ is the statistical operator. Using last expression
it is easy to present Green's function generating functional by the
following expression
\begin{equation}
Z[j]=<{\rm T}\exp \left[i\int d^4 y j(y)\hat{\phi}(y) \right]>=
{\rm Tr} \left\{\hat{\rho}
{\rm T}\exp \left[i\int d^4 y j(y)\hat{\phi}(y) \right] \right\}.
\label{3}
\end{equation}
 From (\ref{3}) an $n$-point Green's function  may be obtained
\begin{equation}
i{\cal G}(x_1,\cdots ,x_n)= \left.
\frac{1}{Z[0]}\frac{\delta}{i\delta j(x_1)}
\cdots \frac{\delta}{i\delta j(x_n)}  Z[j] \right|_{j=0}.
\label{4}
\end{equation}
These functions are investigated in every details for the equilibrium case
\cite{NS,UMT,JK,LvW,RR} when
\begin{equation}
\hat{\rho}=\hat{\rho}_0=Q_{0}^{-1}e^{-\beta _0\hat{H}},
\label{5}
\end{equation}
with $\beta_0=1/T_0$ being the inverse temperature, $\hat{H}$ ---
Hamiltonian and $Q_0$ --- partition function of the system. In the
nonequilibrium case I shall use NSO introduced by Zubarev \cite{Zub1,Zub}
\begin{equation}
\hat{\rho}_{\varepsilon}(t)= \varepsilon\int_{-\infty}^{0} d t_1
e^{\varepsilon t_1}Q_{\varepsilon}^{-1}(t)
\exp \left[- \int d^3x \beta^{\mu}
({\bf x},t+t_1)\hat{T}_{0 \mu} ({\bf x},-T+t_1)\right],
\label{6}
\end{equation}
where $\hat{T}_{\mu \nu}$ is the energy-momentum tensor and
$\beta^{\mu}(x)=\beta(x)u^{\mu}(x)$.  Here $u^{\mu}(x)$ is the local
hydrodynamical velocity and $\beta(x)$ is an inverse local temperature
$\beta(x)=1/T(x)$.

Now consider the inverse temperature distribution slightly differed from the
certain constant value $\beta_{0}^{\mu}=\beta_{0}u^{\mu}$ by a small
parameter
\begin{equation}
\beta_{1}^{\mu}(x)=\beta^{\mu}(x)-\beta_{0}^{\mu}.
\label{8}
\end{equation}
In the rest frame (where $u_{0}^{\mu}=(1,0,0,0)$), NSO has the following
form:
\begin{eqnarray}
\hat{\rho}_{\varepsilon}(t)= \varepsilon\int_{-\infty}^{0}
d t_1 e^{\varepsilon t_1}
Q_{\varepsilon}^{-1}(t)\exp \left[-\beta _0 \hat{H}- \right.
\; \; \;  \nonumber \\ \left.
- \int d^3 x \beta^{\mu}_1 ({\bf x},t+t_1)
\hat{T}_{0 \mu} ({\bf x},-T+t_1) \right].
\label{9}
\end{eqnarray}
It is possible now to rewrite NSO (\ref{6}) as a product of the equilibrium
statistical operator (\ref{5}) and the certain residual.  Let us use the
well-known formula \cite{FKS}
\begin{equation}
\exp\left[- \beta_0 \hat{H}+\hat{B}(-T)\right]=
e^{- \beta_0 \hat{H}} {\rm T}_{\tau} \left\{\exp \left[i \int_{0}^{-i\beta}
d \tau \beta_{0}^{-1}\hat{B}(\tau) \right] \right\},
\label{10}
\end{equation}
where $T_{\tau}$ is ordering over $\tau$-variable along the segment of
integration and with
\begin{equation}
\hat{B}(z_2)=e^{i(z_2-z_1)\hat{H}}\hat{B}(z_1)e^{-i(z_2-z_1)\hat{H}}
\label{10a}
\end{equation}
continued to the complex values of $z$ variable. Integration over
$\tau$ in (\ref{10}) may be generalised \cite{BP} on the well-known in
the real-time equilibrium theories contour $C$ (fig.~1) \cite{M,LvW} to
get
\begin{equation}
\exp\left[- \beta_0 \hat{H}+\hat{B}(-T)\right]=
e^{- \beta_0 \hat{H}} {\rm T}_C \left\{\exp \left[i \int_{C} d x_0
\beta_{0}^{-1}\hat{B}(x_0) \right] \right\},
\label{11}
\end{equation}
where ${\rm T}_C$ --- is an ordering by the time variable along $C$.
Now for the NSO we get
\begin{eqnarray}
\hat{\rho}_{\varepsilon}(t)=\varepsilon \int_{-\infty}^{0}
d t_1 e^{\varepsilon t_1} Q_{\varepsilon}^{-1}(t)
e^{-\beta _0 \hat{H}} \times \; \; \; \; \nonumber \\
\times   {\rm T}_C \left\{\exp \left[
-i \int_{C} d^4 x \frac{\beta^{\mu}_1 ({\bf x},t+t_1)}{\beta_0}
\hat{T}_{0 \mu} ({\bf x},x_0+t_1) \right] \right\},
\label{11b}
\end{eqnarray}
where ${\rm T}_C$ is ordering only over $x_0$ and $d^4x=dx_0d^3x$.
However we shall consider ${\rm T}_C$ as an ordering over a whole
time-argument $x_0+t_1$ of operators (it is possible since $t_1$  is a
constant shift for ${\rm T}_C$ ordering and does not change the time order
of the operators).

Expression (\ref{11b}) is the expected perturbative expansion of the NSO in
a series of time-ordered operator products which take into account effects
that are nonlinear in inverse temperature variations $\beta_1^{\mu}$.  It
reduces the calculation of a nonequilibrium average
\begin{equation}
{\rm Tr}\left(\hat{\rho}_{\varepsilon}{\rm T}e^{j\hat{\phi}}\right),
\label{12}
\end{equation}
where
\begin{equation}
e^{j\hat{\phi}}=\exp \left[i\int d^4yj(y)\hat{\phi}(y)\right].
\label{A3}
\end{equation}
to the calculation of a series of equilibrium averages like
\begin{equation}
{\rm Tr}\left(\hat{\rho}_{0}{\rm T}_C\left\{(\beta_1^{\mu}\hat{T}_{0
\mu})\right\}^n {\rm  T}e^{j\hat{\phi}}\right),
\label{13}
\end{equation}
where $\beta_1^{\mu}\hat{T}_{0 \mu}$ is notation for the integral in the last
exponent
in the (\ref{11b}).  Note that a distinctive feature of this expansion is that
the small parameter $\beta_1^{\mu}$ being not from the Hamiltonian $\hat{H}$.
Now, an ordinary expansion over parameters from $\hat{H}$ is needed.

\section{Calculation of the double-time or\-dered op\-erator averages in
vacuum}
\label{Ca}
Let us carry out a perturbative calculation in the vacuum quantum field
theory of an average like \begin{equation}
G(x_1,x_2,x_3,x_4)=<0\left|{\rm T}
\left[\hat{\phi}(t_1)\hat{\phi}(t_2)\right]
{\rm T}\left[\hat{\phi}(t_3)\hat{\phi}(t_4)\right]
\right|0>
\label{A1}
\end{equation}
We are based on the real-time method, introduced in the thermo-field dynamics
and presented in terms of path integrals \cite{NS} but apply it in the vacuum
case.  We can use the fact that deformation of a time-integration contour makes
possible to calculate not only time ordered, but non-ordered and
antitemporal-ordered products of field operators in terms of path integrals.
We
shall show here, that addition to the time-integration axis permits to
calculate
vacuum averages of the double time-ordered and mixed ordered products of a
field
operators in terms of path integrals.

Function (\ref{A1}) may be obtained from the generating functional
\begin{equation}
{\cal Z}[j,j']=<0\left| {\rm T} e^{j\hat{\phi}}{\rm T}e^{j'\hat{\phi}'}
\right|0>,
\label{A2}
\end{equation}
where $e^{j\hat{\phi}}$ is given in (\ref{A3}). Indeed,
\begin{equation}
G(x_1,x_2,x_3,x_4)=\left.
\frac{1}{{\cal Z}[0]}\frac{\delta}{i\delta j(t_1)}
\frac{\delta}{i\delta j(t_2)}\frac{\delta}{i\delta j'(t_3)}
\frac{\delta}{i\delta j'(t_4)}{\cal Z}[j,j'] \right|_{j,j'=0}.
\label{A4}
\end{equation}
To present (\ref{A2}) in terms of path integrals it is important to
reduce multi time-ordered product to once ordered one. It is easy
to do by means of a transformation of the time-integration contour
from real axis to the contour ${\cal C}$ (Fig.~2) on the complex time
$t$-plane. Now the generating functional is
\begin{eqnarray}
{\cal Z}[j]=<0 \left|{\rm T}_{{\cal C}}e^{j\phi} \right| 0>=
\int {\cal D}_{{\cal C}} \phi \exp \left[ i\int_{{\cal C}}d^4 x
\left( {\cal L}(\phi)+j(x) \phi (x) \right)\right]= \nonumber \\
=\exp \left[-i\int_{{\cal C}}d^4x V\left[\frac{\delta}{i\delta j(x)}\right]
\right] \times\label{A5}\\ \times
\exp \left[ -\frac{1}{2} \int_{{\cal C}}d^4 z_1 d^4 z_2 j(z_1)
{\cal D} (z_1-z_2) j(z_2) \right].
\nonumber
\end{eqnarray}
Let us ascribe to the various segments of ${\cal C}$ indices 1,2 and 3.
Then (\ref{A5}) may be presented in the form
\begin{eqnarray}
{\cal Z}[j_{1,2,3}]=\exp\left[-i\int\left(
V\left[\frac{\delta}{i\delta j_1}\right]-
V\left[\frac{\delta}{i\delta j_2}\right]+
V\left[\frac{\delta}{i\delta j_3}\right]\right)\right]\times \nonumber \\
\times \exp\left[-\frac{i}{2}\int j_a{\cal D}_{ab}j_b\right],
\label{A6}
\end{eqnarray}
where the propagator ${\cal D}_{ab}$ in the $\lambda\phi^4$-theory is
\begin{equation}
i{\cal D}_{ab}=\left[
\matrix{
{\cal D}_F  & {\cal D}^{>}  & {\cal D}^{>} \cr
{\cal D}^{<}  & {\cal D}_F^{\ast}  & {\cal D}^{>} \cr
{\cal D}^{<}  & {\cal D}^{<}  & {\cal D}_F \cr
}
\right]
\label{A7}
\end{equation}
\begin{equation}
{\cal D}_F=\frac{i}{p^2-m^2+i\varepsilon},{\ }{\cal D}_F^{\ast}=
\frac{-i}{p^2-m^2-i\varepsilon}
\label{A8}
\end{equation}
\begin{equation}
{\cal D}^{>}=-2\pi\theta(p_0)\delta(p^2-m^2),{\ }
{\cal D}^{<}=-2\pi\theta(-p_0)\delta(p^2-m^2).
\label{A88}
\end{equation}
Function (\ref{A1}) may be obtained from (\ref{A6}) analogical to the
expression (\ref{A4}), where $j=j_1$, $j'=j_3$. Generating functional
(\ref{A6}) makes it also possible to calculate expression with an
antitemporal order hence segment ${\cal C}_2$  of ${\cal C}$ (Fig.~2) is
the path from $+\infty $ to $-\infty $. For example
\begin{eqnarray}
G(x_1,x_2,x_3,x_4)=<0\left|{\rm T}
\left[ \phi(t_1)\phi(t_2)\right]
\tilde{{\rm T}}\left[ \phi(t_3)\phi(t_4)\right]
\right|0>=\nonumber \\
=\left.
\frac{1}{{\cal Z}[0]}\frac{\delta}{i\delta j_{1}(t_1)}
\frac{\delta}{i\delta j_{1}(t_2)}\frac{\delta}{i\delta j_{2}(t_3)}
\frac{\delta}{i\delta j_{2}(t_4)}{\cal Z}[j_{1,2,3}] \right|_{j_{1,2,3}=0},
\label{A9}
\end{eqnarray}
where $\tilde{{\rm T}}$ denotes an antitemporal ordering of operators.

This can be easily generalised to every number of ordered and
anti-ordered groups of operators after introducing a more
complicated contour (see fig.~3), containing corresponding number
of segments.

\section{Nonequilibrium generating functional}

Let us return to the generating functional  (\ref{3}) with $\hat{\rho}$  in
the NSO form (\ref{6}), decomposed according to the (\ref{11b}). As it was
mentioned at the end of the Sec. \ref{Ho}, this decomposition leads to the
equilibrium calculation of the double time-ordered product (\ref{13}):
\begin{eqnarray}
T_C \left\{\exp \left[
-i \int_{C} d^4 x \frac{\beta^{\mu}_{1}({\bf x},t+t_1)}{\beta_0}
\hat{T}_{0 \mu} ({\bf x},x_0+t_1) \right] \right\}
\times \nonumber \\ \times
T \left\{ \exp \left[i\int d^4 y j(y)\hat{\phi}(y)\right] \right\}.
\label{G1}
\end{eqnarray}
To calculate (\ref{G1}) the following generating functional may be
introduced
\begin{eqnarray}
Z[j,j']={\rm Tr} \left\{\hat{\rho_0}
T_C \exp\left[i\int d^4 y' j(y')\hat{\phi}(y') \right]
T\exp \left[ i\int d^4 y j(y)\hat{\phi}(y) \right] \right\}.
\label{G2}
\end{eqnarray}
Using identity
\begin{equation}
F(\phi)e^{j\phi}=F\left[\frac{\delta}{i\delta j}\right]e^{j\phi}
\label{id}
\end{equation}
we may present  nonequilibrium generating functional in terms of (\ref{G2})
\begin{eqnarray}
{\cal Z}[j]=
\varepsilon \int_{-\infty}^{0} d t_1 e^{\varepsilon t_1}
\times \nonumber \\ \times \left.
\exp \left[-i \int_{C} d^4 x \frac{\beta^{\mu}_1 ({\bf x},t+t_1)}{\beta_0}
T_{0 \mu}\left[\frac{\delta}{i\delta j({\bf x},x_0+t_1)}\right]
\right] Z[j,j'] \right|_{j'=0},
\label{G3}
\end{eqnarray}
where $T_{\mu \nu}[\delta/i\delta j]$ is the same function of
$\delta/i\delta j$ as $T_{\mu \nu}(\phi)$ of $\phi$.

As it was done in the previous Sec., we introduce now the contour $\cal C$
on fig.~4.

In terms of  ${\cal C}$, (\ref{G3}) may be presented in the form
\begin{eqnarray}
{\cal Z}[j]=\lim_{\varepsilon \rightarrow +0} \varepsilon\int_{-\infty}^{0}
dt_1 e^{\varepsilon t_1}\times  \nonumber \\ \times
 \exp \left[-i \int_{{\cal C}_{3,4,5}}
d^4x \frac{\beta^{\mu}_1 ({\bf x},t+t_1)}{\beta_0}
T_{0 \mu} \left[\frac{\delta}{i\delta j({\bf x},x_0+t_1)}\right] \right]
Z[j],
\label{G4}
\end{eqnarray}
where
\begin{equation}
Z[j]={\rm Tr} \left\{e^{- \beta_0 \hat{H}}
T_{\cal C} \exp \left[i\int_{\cal C} d^4 y j(y)\hat{\phi}(y)
\right] \right\}.
\label{G5}
\end{equation}
Now we may present (\ref{G5}) in terms of path integrals \cite{B}.
\begin{eqnarray}
Z[j]=\int{\cal D}_{{\cal C}} \phi \exp \left[ i\int_{{\cal C}}d^4 x
\left( {\cal L}(\phi)+j(x) \phi (x) \right)\right]=
\nonumber \\
=\exp \left\{-i\int_{{\cal C}}d^4x V\left[\frac{\delta}{i\delta j}
\right]\right\}
\exp \left[ -\frac{1}{2} \int_{{\cal C}}d^4 z_1 d^4 z_2 j(z_1)
{\cal D}_{\beta_0} (z_1-z_2) j(z_2) \right]
\label{FI}
\end{eqnarray}
with the periodicity condition for the field \cite{NS,LvW,RR,M}
\begin{equation}
\phi (-T)= \phi (-T-i \beta_0).
\label{KMS}
\end{equation}
Here ${\cal D}_{\beta_0}(x-y)$ is a free Green function on
the contour ${\cal C}$ and  $V$ --- is the interacting part of the
Lagrangian.

Contrary to the vacuum theory, this one contains  propagators of
$4\times 4$ dimension \cite{B}.

\section{Ghost fields}

Let us mention the role of the ghost fields \cite{NS}. It was proved by
Dolan and Jackiw \cite{DJ} that real-time QFT at  finite temperature
obtained from the vacuum QFT only with free Green's functions replaced
by the thermal free Green's functions suffer  from pathological pinch
singularities \cite{IZ}. Correct real-time  quantum field theory was
introduced by Mills \cite{M} and it is obtained from the density
matrix. An introduction of the more complicate time contour is required
by the density matrix. This contour contains real-time axis that is
time integration path in the vacuum QFT, as first segment of the whole
time integration path. Fields defined on additional parts of time path
are called ghost fields.  They occur only in the internal lines after
perturbative decomposition and are inevitable for the cancellation of
the pinch singularities in the  multi-loop perturbative expansion
\cite{NS,LvW}. This singularities may occur in the product of $\delta$
- functions from (\ref{A88}) \cite{DJ}.

Thus, correct real-time QFT at finite temperatures may be obtained from
the vacuum QFT with free Green's functions replaced by the thermal ones
and completed by the Green functions of thermal ghosts \cite{NS} at the
internal lines.

To calculate double-time ordered operator product in vacuum we also
have changed time integration contour. But it consists of to segments
where external fields are defined. So, field functions $\phi_1$  and
$\phi_3$  are real and $\phi_2$ ---  ghost fields in the equation
(\ref{A9}). Field $\phi_2$, as it stems from (\ref{A6}) may occur only
on internal lines of the diagrams in the perturbative calculation of an
eq.(\ref{A1}). Their existence again is very important for the
cancellation of the ``pinch singularities''.

At the end, it is important to emphasize that real-time thermal theory
without ghosts \cite{DJ} or with ghosts defined by hands \cite{AS}
contains singularities. The best way to avoid singularity is to
construct a theory on the basis of density matrix or other fundamental
principles.

\vspace{0.2in}
{\Large \bf Acknowledgements}
\vspace{0.2in}

I am thankful to J.~D.~Manjavidze, I.~Paziashvili and E.~K.~Sarkisyan for
interesting conversations.  I would like to thank also T.~Kibble and P.~Goddard
for the hospitality at the I.Newton Institute at The Cambridge University,
where
this paper was discussed after it was finished.  This work was supported in
part
by G.~Soros foundation through the APS and ISF.

\end{document}